\documentclass[12pt]{article}

\def\BEq{\begin{equation}}
\def\EEq{\end{equation}}
\def\BEqA{\begin{eqnarray}}
\def\EEqA{\end{eqnarray}}
\def\BEn{\begin{enumerate}}
\def\EEn{\end{enumerate}}

\usepackage{makeidx,graphics,latexsym}

\begin{document}

\title{{\bf Controlled-NOT gate design for Josephson phase qubits with 
tunable inductive coupling: Weyl chamber steering and area theorem}}

\author{
Andrei
Galiautdinov and Michael Geller
\\
\normalsize{\it Department of Physics and Astronomy,
University of Georgia, Athens, Georgia
30602, U.S.A.}
}

\date{\today}

\maketitle

\abstract{
Superconducting qubits with tunable coupling are ideally suited for fast 
and accurate implementation
of quantum logic. Here we present a simple approach, based on Weyl chamber 
steering, to CNOT 
gate design for inductively coupled phase 
qubits with tunable coupling strength $g$. In the presence of simultaneous 
rf pulses 
on the individual qubits that appropriately {\it track} the coupling strength 
as it is varied, 
we show that an infinite family of switching sequences preserving the time 
integral or ``area"
of $g$ can be used to generate CNOT logic. We demonstrate our approach by 
considering time-dependencies most likely to be used
in actual implementations: trapezoidal, sine, and 
soft quartic (also known as Landau's hat).}

\vskip10pt
PACS number(s): 03.67.Lx, 85.25.Cp
\vskip10pt

Superconducting circuits containing Josephson junctions are promising 
candidates for scalable solid-state quantum computing architectures 
\cite{MakhlinRMP01,YouPT05, YuSci02,MartinisPRL02,BerkleySci03,McDermottSci05,
SteffenSci06}. 
In this paper we
describe a pulse switching design suitable for generation of high-fidelity 
CNOT logic by systems with tunable inductive coupling. Such a tunable coupling 
has been recently  
demonstrated
for flux qubits \cite{CLARKESCI}.

The Hamiltonian (in the doubly rotating frame) for resonant phase qubits is 
\cite{MYCNOTPAPER1}
\begin{equation}
H = \sum_i \frac{\Omega_i(t)}{2}\sigma^x_i + \frac{g(t)}{2} \left(\sigma_1^x 
\sigma_2^x + \sigma_1^y \sigma_2^y + k \sigma_1^z \sigma_2^z\right),
\label{rwa hamiltonian}
\end{equation}
where we have emphasized the fact that both the Rabi frequencies $\Omega_i$ 
and the qubit-qubit interaction strength $g$ are time dependent. Here $k$ is 
a constant (real) parameter of order unity that depends on the qubit flux bias. 
The Hamiltonian (\ref{rwa hamiltonian}) neglects rapidly oscillating terms with 
vanishing time-averages (i.e., we use the rotating-wave approximation). When $k=0$, 
the Hamiltonian (\ref{rwa hamiltonian}) also describes phase qubits with tunable 
capacitive coupling.

Our gate construction relies on the identity
\begin{equation}
e^{-i \frac{\pi}{4}(\Lambda_1 \sigma^x_1 + \Lambda_2 \sigma^x_2 +
\sigma_1^x \sigma_2^x + \sigma_1^y \sigma_2^y + k \sigma_1^z \sigma_2^z)}
= {\rm CNOT_{Weyl}}
\label{steering identity}
\end{equation}
derived in Ref. \cite{MYCNOTPAPER1}, where 
\BEq
\Lambda_{1,2}(k) = \sqrt{16-\bigg(\frac{k-1}{2}\bigg)^2} \pm 
\sqrt{16-\bigg(\frac{k+1}{2}\bigg)^2} ,
\label{Lambda definitions}
\EEq
and
\begin{equation}
{\rm CNOT_{Weyl}} \equiv e^{-i \frac{\pi}{4} \sigma_1^x \sigma_2^x}
= 
\frac{1}{\sqrt{2}}
\left[
\begin{array}{cccc}
1&0&0&-i\\
0&1&-i&0\\
0&-i&1&0\\
-i&0&0&1\\
\end{array}
\right].
\end{equation}
Here ${\rm CNOT_{Weyl}}$ is the gate in $SU(4)$ local equivalence class 
of canonical CNOT \cite{MAKHLIN,ZHANG, ZHANG1},
\begin{equation}
{\rm CNOT} \equiv 
\left[
\begin{array}{cccc}
1&0&0&0\\
0&1&0&0\\
0&0&0&1\\
0&0&1&0\\
\end{array}
\right],
\end{equation}
which belongs to the Weyl chamber \cite{weylchambernote}. The expression 
for the $\Lambda_i(k)$ given here is valid for $-7 \leq k \leq 7$; for 
expressions valid for larger values of $|k|$ see Ref. \cite{MYCNOTPAPER1}. 
The CNOT can be generated (up to an overall phase factor) from 
${\rm CNOT_{Weyl}}$ by applying local $SU(2) \times SU(2)$ rotations,
\begin{equation}
{\rm CNOT} = e^{i \frac{\pi}{4}} \, e^{-i\frac{\pi}{4}\, \sigma^y_1} \, 
e^{i\frac{\pi}{4}\left(\sigma^x_1 - \sigma^x_2 \right)} \, {\rm CNOT_{Weyl}} \, 
e^{i\frac{\pi}{4}\,\sigma^y_1}.
\label{CNOT identity}
\end{equation}
The local rotations in (\ref{CNOT identity}) are to be performed with $g=0$.

An ``area" theorem follows by varying the Rabi frequencies to {\it track} the 
time-dependent coupling strength,
\begin{equation}
\Omega_i(t) = \Lambda_i \, g(t),
\label{tracking omegas}
\end{equation}
with the fixed constants $\Lambda_i$ given above. Then (\ref{rwa hamiltonian}) 
becomes
\begin{equation}
H= g(t) \, {\cal H},
\label{tracking hamiltonian}
\end{equation}
with
\begin{equation}
{\cal H} \equiv \frac{\Lambda_1 \sigma^x_1 + \Lambda_2 \sigma^x_2 +
\sigma_1^x \sigma_2^x + \sigma_1^y \sigma_2^y + k \sigma_1^z \sigma_2^z}{2}
\end{equation} 
a fixed matrix. The Hamiltonian (\ref{tracking hamiltonian}) commutes with 
itself at different times, so the 
time-evolution operator is
\begin{equation}
U = e^{-i \theta {\cal H}},
\end{equation}
where 
\begin{equation}
\theta \equiv \frac{1}{\hbar} \int_0^{t_1} dt \, g(t).
\label{evolution angle}
\end{equation}
Here we have assumed that the interaction is turned on at time $t=0$ and is 
turned off at some later time $t_1$. If we choose the angle (\ref{evolution 
angle}) to be $\theta = \pi/2$, the identity (\ref{steering identity}) 
shows that we can construct a CNOT gate.

Our CNOT gate implementation thus proceeds as follows: 

\begin{enumerate}

\item First the coupling is turned off and the local rotation 
$R_y(-\frac{\pi}{2})_1$ on qubit 1 is performed. 

\item Then $g$ is turned on and off according to some experimentally 
convenient switching profile, such that
\begin{equation}
 \int_0^{t_1} dt \, g(t) = \frac{\pi \hbar}{2},
\label{area identity}
\end{equation}
with $\Omega_i(t)$ tracking it in accordance with (\ref{tracking omegas})

\item Simultaneous rotations $R_x(-\frac{\pi}{2})_1 \otimes R_x(\frac{\pi}{2})_2$ 
are
applied to the qubits with $g=0$.

\item Finally, a rotation $R_y(\frac{\pi}{2})_1$ is applied to qubit 1 with 
the coupling off.
\end{enumerate}

We turn now to a discussion of three switching profile examples.

\begin{description}

\item{\it Trapezoidal switching} --- Any trapezoidal pulse \cite{WELLCNOT, PLOURDE} 
can be broken down into three parts:

\begin{enumerate}
\item Tuning with
\BEq
\label{eq:Htune}
g(t) = \frac{gt}{\epsilon t_1},
\EEq
where $0\leq \epsilon \leq 1/2$ 
characterizes the ramping fraction of the total
time $t_1$; 

\item Evolution with resonant Hamiltonian $H = g{\cal H}$ for $t=(1-2\epsilon)t_1$;

\item Detuning with 
\BEq
\label{eq:Hdetune}
g(t) = 
g\left( 1- \frac{t}{\epsilon t_1}\right).
\EEq
\end{enumerate}
Then
\BEq
\label{eq:FULLGATE}
\theta =  \frac{gt_1}{\hbar}\,(1-\epsilon).
\EEq

We can smooth out the upper trapezoidal corners by considering inverted quadratic, 
quartic, and other higher order pulses with time-dependent prefactors of the form
\BEq
\label{eq:Hquart}
g_{2n}(t) = g\left[1 - 2^{2n}\left(\frac{t}{t_1}-\frac{1}{2}\right)^{2n}\right], 
\quad n=1,2,3,\dots .
\EEq
This leads to 
\BEq
\theta = \frac{g t_1}{\hbar}  
\,\left[
1-\frac{1}{2n+1}
\right].
\EEq

\item{\it Sinusoidal switching} --- In this case the time dependence is
\BEq
\label{eq:Hsin}
g(t) = 
\frac{g}{2}\left[1-\cos\left(\frac{2\pi t}{t_1}\right)\right],
\EEq
and
\BEq
\label{eq:sinPULSE}
\theta = \frac{g t_1}{2\hbar}.
\EEq

\item{\it Landau's hat} --- This pulse is in the form of the ``middle part'' of the 
famous curve used by Landau in his 
theory of phase transitions. It is mathematically simple, soft, and faster than 
sinusoidal. The time dependence is
\BEq
\label{eq:Hlandau}
g(t) = 
g \left(1 + \left(2t/t_1-1\right)^4
-2\left(2t/t_1-1\right)^2\right),
\EEq
giving
\BEq
\label{eq:quartPULSE}
\theta = \frac{8g t_1}{15\hbar}.
\EEq
\end{description}
The corresponding gate times $t_1$ can now be found using (\ref{evolution angle}) 
and (\ref{area identity}):
\begin{equation}
\label{eq:SUMMARYTABLE}
 \begin{array}{|l|l|l|}\hline
 {\rm \bf Switching\; mechanism}	&
 {\rm \bf Pulse\; profile} &
 t_1, \times \pi\hbar/(2g)   \\ \hline
\epsilon=0.0000\; {\rm (none)}	
&  {\rm rectangular}  & 1.0000	\\
\epsilon=0.0250\; {\rm (fast)}	&  {\rm trapezoidal}  & 1.0256		\\
\epsilon=0.2000\; {\rm (moderate)}	&  {\rm trapezoidal}  &1.2500		\\
\epsilon=0.5000\; {\rm (slow)}	&  {\rm triangular} & 2.0000	\\
\hline
n=1 & {\rm inverted\; quadratic}& 1.5000 \\
n=2 & {\rm inverted\; quartic}& 1.2500 \\
n=3 & {\rm inverted\; hexagonic}& 1.1667 \\
n=4 & {\rm inverted\; octagonic} & 1.1250	\\
\hline
{\rm sinusoidal}	&  {\rm inverted\; cosine} & 2.0000	\\
\hline
{\rm soft\; quartic}	&  {\rm Landau's\; hat} & 1.8750	\\
\hline
 \end{array}
\end{equation}

In summary, we have shown how to implement a CNOT gate for phase
qubits with tunable inductive coupling using a construction based on
Weyl chamber steering. Two approximations have been made in our
analysis, the rotating wave approximation and the neglect of leakage
to higher lying (non-qubit) states.


\section*{Acknowledgments}

This work was supported by the Disruptive Technology Office under
grant W911NF-04-1-0204 and by the National Science Foundation under
grant CMS-0404031.

The authors would like to thank Emily Pritchett 
and Andrew Sornborger for helpful discussions.


\begin{thebibliography}{}
\bibitem{weylchambernote}
We use a Weyl chamber defined slightly differently than that of Ref. \cite{ZHANG}; 
the class vectors are related by an overall minus sign.
\end{thebibliography}
\end{document}